%
\let\useblackboard=\iftrue
%
%
\newfam\black
\input harvmac.tex
\def\Title#1#2{\rightline{#1}
\ifx\answ\bigans\nopagenumbers\pageno0\vskip1in%
\baselineskip 15pt plus 1pt minus 1pt
\else
\def\listrefs{\footatend\vskip 1in\immediate\closeout\rfile\writestoppt
\baselineskip=14pt\centerline{{\bf References}}\bigskip{\frenchspacing%
\parindent=20pt\escapechar=` \input
refs.tmp\vfill\eject}\nonfrenchspacing}
\pageno1\vskip.8in\fi \centerline{\titlefont #2}\vskip .5in}

\ifx\answ\bigans\def\tcbreak#1{}\else\def\tcbreak#1{\cr&{#1}}\fi
\useblackboard
\message{If you do not have msbm (blackboard bold) fonts,}
\message{change the option at the top of the tex file.}
\font\blackboard=msbm10 scaled \magstep1
\font\blackboards=msbm7
\font\blackboardss=msbm5
\textfont\black=\blackboard
\scriptfont\black=\blackboards
\scriptscriptfont\black=\blackboardss

\else

\fi
%
\def\yboxit#1#2{\vbox{\hrule height #1 \hbox{\vrule width #1
\vbox{#2}\vrule width #1 }\hrule height #1 }}
\def\fillbox#1{\hbox to #1{\vbox to #1{\vfil}\hfil}}
\def\ybox{{\lower 1.3pt \yboxit{0.4pt}{\fillbox{8pt}}\hskip-0.2pt}}

\def\comments#1{}

\def\II{\relax{I\kern-.07em I}}

\def\IZ{\relax\ifmmode\mathchoice
{\hbox{\cmss Z\kern-.4em Z}}{\hbox{\cmss Z\kern-.4em Z}}
{\lower.9pt\hbox{\cmsss Z\kern-.4em Z}}
{\lower1.2pt\hbox{\cmsss Z\kern-.4em Z}}\else{\cmss Z\kern-.4em
Z}\fi}
\def\IB{\relax{\rm I\kern-.18em B}}
\def\IC{{\relax\hbox{$\inbar\kern-.3em{\rm C}$}}}
\def\ID{\relax{\rm I\kern-.18em D}}
\def\IE{\relax{\rm I\kern-.18em E}}
\def\IF{\relax{\rm I\kern-.18em F}}
\def\IG{\relax\hbox{$\inbar\kern-.3em{\rm G}$}}
\def\IGa{\relax\hbox{${\rm I}\kern-.18em\Gamma$}}
\def\IH{\relax{\rm I\kern-.18em H}}
\def\II{\relax{\rm I\kern-.18em I}}
\def\IK{\relax{\rm I\kern-.18em K}}
\def\IP{\relax{\rm I\kern-.18em P}}

\font\cmss=cmss10 \font\cmsss=cmss10 at 7pt
\def\IR{\relax{\rm I\kern-.18em R}}

\def\IZ{\relax\ifmmode\mathchoice
{\hbox{\cmss Z\kern-.4em Z}}{\hbox{\cmss Z\kern-.4em Z}}
{\lower.9pt\hbox{\cmsss Z\kern-.4em Z}}
{\lower1.2pt\hbox{\cmsss Z\kern-.4em Z}}\else{\cmss Z\kern-.4em
Z}\fi}
\def\IB{\relax{\rm I\kern-.18em B}}
\def\IC{{\relax\hbox{$\inbar\kern-.3em{\rm C}$}}}
\def\ID{\relax{\rm I\kern-.18em D}}
\def\IE{\relax{\rm I\kern-.18em E}}
\def\IF{\relax{\rm I\kern-.18em F}}
\def\IG{\relax\hbox{$\inbar\kern-.3em{\rm G}$}}
\def\IGa{\relax\hbox{${\rm I}\kern-.18em\Gamma$}}
\def\IH{\relax{\rm I\kern-.18em H}}
\def\II{\relax{\rm I\kern-.18em I}}
\def\IK{\relax{\rm I\kern-.18em K}}
\def\IP{\relax{\rm I\kern-.18em P}}

\font\cmss=cmss10 \font\cmsss=cmss10 at 7pt
\def\IR{\relax{\rm I\kern-.18em R}}

\def\tilde{\widetilde}
\def\frac#1#2{{{#1} \over {#2}}}

\def\npb#1#2#3{{\it Nucl. Phys.} {\bf B#1,} #2 (19#3)}

\def\plb#1#2#3{{\it Phys. Lett.} {\bf B#1,} #2 (19#3)}

\def\prd#1#2#3{{\it Phys. Rev.} {\bf D#1,} #2 (19#3)}

\Title{ \vbox{\baselineskip12pt\hbox{hep-th/9704089}
\hbox{RU-97-23, UTTG-12-97}}}
{\vbox{\centerline{Matrix Description of M-theory on $T^4$ and $T^5$}}}

\centerline{Micha Berkooz${}^1$, Moshe Rozali${}^2$ and Nathan
Seiberg${}^1$}
\smallskip
\smallskip
\centerline{${}^1$ Department of Physics and Astronomy}
\centerline{Rutgers University }
\centerline{Piscataway, NJ 08855-0849}
\smallskip
\smallskip
\centerline{${}^2$ Theory Group,  Department of Physics}
\centerline{University of Texas}
\centerline{Austin, Texas 78712}

\bigskip
\bigskip
\noindent
We study the Matrix theory description of M-theory compactified on
$T^4$ and $T^5$. M-theory on $T^4$ is described by the six dimensional
(2,0) fixed point field theory compactified on a five torus,
$\widetilde T^5$.  For M-theory on $T^5$ we suggest the existence of a
new theory which is compactified on a $\widetilde T^5$.  The IR
description of this theory is given by the (2,0) theory with a
compactified moduli space.  This new theory appears to be a new kind
of a non-critical string theory.  Clearly, these two descriptions
differ from the ``standard'' Super-Yang-Mills on the dual torus
prescription.

\Date{May 1997}

\newsec{Introduction}

In the last few months a significant amount of evidence has
accumulated in support of the conjecture of Banks, Fischler, Shenker
and Susskind on the non-perturbative formulation of M-theory
\ref\bfss{T. Banks, W. Fischler, S. Shenker and L. Susskind,
hep-th/9610043, \prd{55}{112}{97}.}.
The structure of 11-dimensional supergravity and the membrane of
M-theory were found in \bfss. The longitudinal 5-brane was found in 
\nref\bd{M. Berkooz and M.R. Douglas, hep-th/9610236,
\plb{395}{196}{97}.}%
\nref\grw{O.J. Ganor, S. Ramgoolam and W. Taylor IV, hep-th/9611202.}%
\nref\bss{T. Banks, N. Seiberg and S.H. Shenker, hep-th/9612157,
\npb{490}{91}{97}.}%
\refs{\bd - \bss}. Various D-branes were constructed in
\bss. Elementary strings and part of their interactions were
constructed in  
\nref\str{L. Motl, hep-th/9701025.}%
\nref\bs{T. Banks and N. Seiberg, hep-th/9702187.}%
\nref\stro{P.-M. Ho and Y.-S. Wu, hep-th/9703016.}%
\nref\strt{R. Dijkgraaf, E. Verlinde and H. Verlinde,
hep-th/9703030.}%
\refs{\str - \strt}.  The interactions of solitonic states
also seem to fit the expected pattern from M-theory
\ref\intr{O. Aharony and M. Berkooz, hep-th/9611215;
G. Lifschytz and Samir D. Mathur, hep-th/9612087;
G. Lifschytz, hep-th/9612223; D. Berenstein and R. Corrado,
hep-th/9702108; G. Lifschytz, hep-th/9703201; I. Cheplev
and A.A. Tseytlin, hep-th/9704127.}.

\nref\tori{W. Taylor IV, hep-th/9611042, \plb{394}{283}{97}.}%
\nref\sus{L. Susskind, hep-th/9611164.}%
\nref\sustwo{S. Sethi and L. Susskind, hep-th/9702101.}%
\nref\torio{D. Berenstein, E. Corrado and J. Distler, hep-th/9704087.}%
\nref\fhrs{W. Fischler, E. Halyo, A. Rajaraman and L. Susskind,
hep-th/9703102.}%

In \refs{\bfss - \bss, \bs, \tori - \fhrs} compactifications on tori
were considered.  It was suggested that M theory on $T^d$ is defined
by $d+1$ dimensional super-Yang-Mills (SYM) on $\hat T^d$ where $\hat
T^d$ is a torus dual to the space-time torus.  However, for $d> 3$
these field theories are not renormalizable, and therefore cannot give
a complete description of the theory.  Furthermore, one cannot define
them as the long distance limit of another theory in $d+1$ dimensions
with the same amount of supersymmetry.  For such a description to
exist, there must be a fixed point of the renormalization group in
$d+1$ dimensions with that supersymmetry.  However, such fixed points
do not exist
\ref\sixteen{N. Seiberg, ``Notes on theories with 16 supercharges,''
RU-97-7, to appear.}.
\nref\nsj{N. Seiberg, talk given at the Jerusalem Winter School on
Strings and Duality, Jan. 1997.}%
\nref\me{M. Rozali, hep-th/9702136.}%

It is possible that these
theories can be defined by a fixed point with fewer supersymmetries.
Another approach, which we will pursue here, is to {\it define} the 4+1
dimensional theory not by a 4+1 dimensional fixed point but by a 5+1
dimensional fixed point with (2,0) supersymmetry \refs{\nsj, \me}.  This
leads to a geometric understanding of the U-duality group
\refs{\me, \fhrs}.  It is important to stress that the 5+1 dimensional
description does not follow from the 4+1 dimensional gauge theory.  It
is a definition of the latter.  We thus propose to define M theory on
$T^4$ as the (2,0) fixed point theory in six dimensions compactified on
$\tilde T^5$.  The precise five torus $\tilde T^5$ and the details of
this construction will be presented in section 2.

Our work has implications to the nature of the underlying space-time.
In the SYM prescription space-time appears as the moduli space of
vacua of the theory reduced to quantum mechanics.  Strictly speaking,
in quantum mechanics the notion of moduli space of vacua is ill
defined.  If the theory has a classical limit, this moduli space is
the target space of the theory in this limit.  In the example of M
theory on $T^4$, the (2,0) theory is inherently quantum mechanical --
the two form whose field strength is self-dual forces $\hbar$ to be
fixed at order one.  Therefore, this theory does not have a classical
limit and hence we cannot define the moduli space of its vacua, when
it is compactified on $\tilde T^5$.  However, if one of the circles of
this $\tilde T^5$ is much smaller than the others, we have an
approximate 4+1 
dimensional SYM with a fixed $\hbar$.  This theory has a well defined
moduli space of vacua.  We see that the notion of space-time becomes
well defined only at various boundaries of the $\tilde T^5$ parameter
space.  We will explain this in more detail at the end of section 2.

For M-theory on $T^5$ we do not have a full description of the
theory. We suggest that it is given by a new theory whose effective IR
description is given in terms of the (2,0) theory with a modified moduli
space, very much like the description of 
\ref\dvv{R. Dijkgraaf, E. Verlinde and H. Verlinde, hep-th/9603126,
\npb{486}{77}{97}; hep-th/9604055, \npb{486}{89}{97}.}, 
although for somewhat different reasons. In the process we identify
all the 16 states that correspond to M-theory wrapped membranes,
wrapped 5-brane and Kaluza-Klein modes.  In a way which is somewhat
reminiscent of the work of \dvv, we also suggest that in this case the
full theory can perhaps be described as a new kind of non-critical
string theory on $\tilde T^5$.  This explains the origin of the
$SO(5,5,\IZ)$ U-duality group as T-duality of this theory.  This will
be discussed in section 3. 

These two proposals present a departure from the SYM prescription of
the Matrix model.  For the theory on $T^4$ we use a field theory which
is not fully understood.  For the theory on $T^5$ our prescription is
clearly incomplete.  For higher dimensional tori we expect new
elements to come in.  In particular, the exceptional U-duality groups
of these theories makes them especially interesting.

\newsec{M Theory on $T^4$}

As a first attempt to define a Matrix Theory description of M-theory
compactified on $T^4$ one studies 4+1 dimensional SYM theory
compactified on the dual torus $\hat T^4$. Its field content is one
vector multiplet in the adjoint of $U(N)$. For simplicity we will take
the torus to be rectangular. M-theory has five dimensionful parameters
which are the Planck length, $l_p$, and the periodicities of the torus,
$L_i$. The parameters of the gauge theory are the dual torus lengths
$\tilde L_i$ and the gauge coupling $g^2$. They are given by
\refs{\me, \fhrs}:
\eqn\sigmadef{\eqalign{
&\tilde L_i= \frac{(2\pi)^2 {l_p}^3}{L_i R} \cr
&g^2= \frac{ (2\pi)^6 {l_p}^6}{L_1 L_2 L_3 L_4 R} \cr}}
where $R$ is the radius of the compactified longitudinal direction.

This gauge theory is not renormalizable and therefore not well defined.
However, it can be used as a low energy effective theory which is valid
at energies below ${1 \over g^2}$.  In this regime several tests of the
description of this theory as a formulation of M-theory were given.
These include the identification of the various 
fluxes in the theory and their relations to the various BPS charges.
Also, the massive ``W-bosons'' of this theory have the right mass (as
a function of the moduli and the radii) to give the proper graviton
scattering.

To define our 4+1 dimensional nonrenormalizable theory we need to give
more information about its short distance degrees of freedom.  On
general grounds, it is impossible to find these UV degrees of freedom
using only the IR information.  In our case, we are guided by the
$SL(5,\IZ)$ U-duality group to suggest that the desired definition of this
theory is in terms of the $(2,0)$ supersymmetric fixed point in six
dimensions.

The $(2,0)$ theory is an interacting quantum field theory at a fixed
point of the renormalization group (for a review, see \sixteen).
It was first discussed in the context of type IIB compactification on a
singular K3
\ref\wittentwoz{E. Witten, hep-th/9507121, Contributed to STRINGS 95:
Future Perspectives in String Theory, Los Angeles, CA, 13-18 Mar 
1995.}
and later in the context of $N$ nearby 5-branes in M-theory
\nref\andy{A. Strominger, hep-th/9512059, \plb{383}{44}{96}.}%
\nref\wittenbranes{E. Witten, hep-th/9512219, \npb{463}{383}{96}.}%
\refs{\andy,\wittenbranes}.  Here we use the same field theory as a
definition of M-theory on $T^4$.  We should stress, to avoid confusion,
that we 
are not defining M theory as the theory of 5-branes.  We simply use
the theory on the 5-brane as an analogue model for the (2,0) field
theory.  It is this field theory which we use.  This distinction will
become more clear below.
 
We thus propose that M-theory on $T^4$ with radii $L_{1,2,3,4}$ is
described by the large $N$ limit of the (2,0) theory
compactified\foot{E. Witten 
\ref\wittenfive{E. Witten,  hep-th/9610234.}
pointed out that this theory needs more data for its definition.  We are
not sure how this affects our proposal.} on a five torus $\tilde T^5$. 
Its five sizes are related to $L_{1,2,3,4}$ in the
following way.  In \sigmadef\ the sizes of the four torus and the
coupling of the 4+1 SYM were given. This SYM has an additional
conserved $U(1)$ current given by $j={}^*(F\wedge F)$. This symmetry
is to be identified with the Kaluza-Klein $U(1)$ symmetry of rotating
around the small circle. The circumference of this circle is $g^2\over
2\pi$.  In the 4+1 SYM an $n$ instanton
configuration, which has energy $4\pi^2n\over g^2$, corresponds to a
state with total momentum $n$ around the small circle and hence we
identify its periodicity with ${g^2\over 2\pi}$.  The 4+1 SYM
description only identifies the charge of this $U(1)$ symmetry.  It
cannot answer questions regarding the detailed nature of these
instantons such as their bound states, etc..  Such questions can be
answered only in a complete description of the short distance theory.

To summarize, we propose that (2,0) theory is compactified on
$\tilde T^5$ whose sizes are given by 
\eqn\tilderdef{\eqalign{
&\tilde L_i= \frac{(2\pi)^2 {l_p}^3}{L_i R} \cr
&\tilde L_5= \frac{ (2\pi)^5 {l_p}^6}{L_1 L_2 L_3 L_4 R}.}}

As a first test of this proposal we point out that at energies much
smaller than ${1 \over g^2} = {1 \over 2\pi \tilde L_5}$ and for $\tilde
L_5 \ll \tilde L_{1,2,3,4}$ our theory 
becomes the 4+1 dimensional SYM.  This follows from the fact that the
dimensional reduction of the two form $B$ whose field strength $H=dB$
is selfdual is a vector in five dimensions.  No scalar emerges from
this dimensional reduction, and therefore the moduli space of vacua of
the compactified (2,0) theory is the same as that of the gauge theory.
We will discuss the details of the moduli space of vacua below.
Furthermore, the excitations of this six dimensional theory include
strings.  As they wind around the circle which brings them to five
dimensions, they yield the W bosons of the 4+1 dimensional SYM.  Their
mass is given correctly by the expression in the (2,0) theory, and
their contribution to the graviton scattering amplitude is as in the
SYM.  However, the 4+1 dimensional SYM is not exact.  Kaluza-Klein
excitations of the 5+1 dimensional theory are also important and lead
to crucial differences between our theory and the 4+1 SYM theory. For
example, the Kaluza-Klein excitations contribute to graviton
scattering terms which are suppressed by the volume of the space-time
$T^4$.  Note that strings which do not wind around the circle are not
new degrees of freedom in the 4+1 SYM.  They are given by classical
solutions (like four dimensional monopoles) in this theory.

A less trivial test of this proposal \me\ is that
this definition of M-theory on $T^4$ makes the U-duality manifest.
The U-duality group in this case is $SL(5,\IZ)$. It is simply the geometric
duality group of $\tilde T^5$.  This symmetry involves mixing the five
radii $\tilde L$ in a way which is complicated as action on the
individual $L_i$.  Since this U-duality group is manifest, so are its
subgroups which appear in compactifications on lower dimensional tori.

We can also compare the parameters of M-theory with the parameters of
this model. In M-theory there is a single dimensionful parameter, $l_p$,
and 14 dimensionless parameters.  These include 10 metric parameters and
4 $C$ field parameters (these include various parameters which we
previously set to zero for simplicity).  In the Matrix theory there seem
to be 15 parameters in the metric of the 5-torus.  However, since the
(2,0) theory is scale invariant, only 14 of them are meaningful -- the
15th sets the scale.  We identify the 14 dimensionless parameters with
the 14 dimensionless parameters of the M-theory compactification.

The various BPS charges which are scalars in the noncompact dimensions
can be identified as fluxes in the 
Matrix theory.  The fluxes in the 4+1 dimensional SYM are 4 electric
fluxes and 6 magnetic ones, all living in the overall $U(1)$ factor of
the $U(N)$ gauge theory. These fluxes correspond to momentum modes of
0-branes on the 4-torus, and membranes wrapped around 2-cycles of the
4-torus. Our 5+1 dimensional theory allows us to identify them as the
10 fluxes $\int dB$ on 3 cycles of our base $\tilde T^5$.  This
expression makes the $SL(5,\IZ)$ transformation properties of the fluxes
manifest.

In the compactified SYM description space-time emerges as the moduli
space of vacua of the quantum mechanical system.  It is easy to see
that no such description is possible in the (2,0) theory.  In fact,
the theory of a single two form $B$, whose field strength $H=dB$ is
self-dual on $\tilde T^5$, has no moduli space of vacua for the $B$
field.  As in the SYM case, we could attempt to construct the light
degrees of freedom on the moduli space by considering the constant
modes of $B$.  The self-duality condition forces these modes to be time
independent and hence the theory has no light modes and no
moduli space of vacua.

In fact, in quantum mechanics (unlike field theory with more than 2
space-time dimensions) there is never a moduli space of vacua.  Only if
the 
theory has a parameter, which can be interpreted as $\hbar$, can we
expect an approximate notion of moduli space of vacua.  Then, for
$\hbar=0$ the classical theory can have many static solutions which we
can identify as its moduli space of vacua, $\cal M$.  When $\hbar $ is
small we can study the full quantum theory by restricting the degrees of
freedom to $\cal M$ and quantizing only them (Born-Oppenheimer
approximation). 

Returning to our (2,0) theory on $\tilde T^5$, we realize that this theory
does not have a free parameter like $\hbar$.  The six dimensional
theory, because of its self-duality, has fixed $\hbar =1$.  Hence, as
we saw above, it cannot have a semiclassical limit with a moduli space
of vacua.  Instead, we can find a moduli space of vacua by creating an
effective $\hbar$.  One way to do that is to consider the limit of
this theory with $\tilde L_5 \ll \tilde L_{1,2,3,4}$.  Then, by going
through the 4+1 dimensional SYM we find a quantum mechanical system
whose moduli space is $T^4$ with sizes $L_{1,2,3,4} \sim {1 \over
\tilde L_{1,2,3,4}}$.  This interpretation of space-time is not
natural, if we pursue it to the region where all the $\tilde L$'s are
of the same order of magnitude.  As any one of the $\tilde L$'s
becomes much smaller than the others there is another natural
interpretation of space time.  This is the essence of U-duality in our
construction.  The natural interpretation of the theory in terms of
space time changes as the five radii $\tilde L$ change.

\newsec{M-Theory on $T^5$}
 
In this section we will attempt to define a Matrix model prescription
for M-theory on $T^5$ with circumferences $L_{1,2,3,4,5}$.  Our proposal
is more speculative than the one 
in the previous section.  Already in the compactification on $T^4$ we
had to depart from the ``SYM on the dual torus'' prescription because
this theory is not renormalizable.  Clearly, for $T^5$ the problem is
even worse.  First, there is no fixed point which flows to SYM in six
dimensions \sixteen.  Furthermore, the description of M-theory on
$T^5$ should include, as one of the radii goes to infinity, the
discussion in the previous section.  Therefore, the theory which is
relevant for M-theory on $T^5$ should be such that it can flow to the
(2,0) theory on $\tilde T^5$ with circumferences
\eqn\tilderdefa{\eqalign{
&\tilde L_i= \frac{(2\pi)^2 {l_p}^3}{L_i R} \cr
&\tilde L_5= \frac{ (2\pi)^5 {l_p}^6}{L_1 L_2 L_3 L_4 R}.}}
This fact shows that the desired theory
cannot be a field theory in more than six dimensions -- in higher than
six dimensions there is no supersymmetry algebra which includes the
six dimensional (2,0) algebra as a subalgebra.

We do not have a complete suggestion for the full theory that
describes M-theory on $T^5$.  Instead, we will present an auxiliary
model that shares the same IR behavior
as this theory.  It is the theory of $N$ M-theoretic 5-branes wrapping 5
cycles of a 6-torus.  The circumferences of the wrapped circles are
$\tilde L_{1,2,3,4,5}$ of \tilderdefa\ and the sixth
circumference  is $L_5$.
This description is particularly useful when the sixth circle is very
large.  It should be stressed that this is merely an analogue model.
The IR behavior of our six dimensional theory is the same as the IR
behavior of this 5-brane theory.  We should not attempt to identify the
base space of our six dimensional theory as the world volume of a
5-brane in M theory. 

One way to motivate this claim is to start with M-theory on $T^5$ with
one of its radii very large, $L_5 \gg L_{1,2,3,4}$.  Then, the
description of M-theory on $T^4$ should be approximately correct.  As in
the previous section, this is given by the (2,0) theory on $\tilde T^5$.
Now, we try to add the effects of the other circle.  We look for a
theory which in the limit $L_5 \rightarrow \infty$ flows to the (2,0)
theory.  Furthermore, its moduli space of vacua should have a circle of
radius $L_5$.  Our proposal in the previous paragraph satisfies these
requirements. 

The analogue model has finite tension
strings which are M-theory membranes stretching from one 5-brane
to another around the compact circle. We expect the full theory to
have these strings also. To see this, consider the following process.
Start with $N$ nearby 5-branes.  They are described by the (2,0) fixed
point.  When they are not on top of each other the spectrum includes
strings.  These originate from membranes which are stretched between
them.  The tension of these strings vanishes as the 5-branes approach
each other.  There are also strings whose tension is
proportional to the circumference of the compact scalar, $L_5$.  They
originate 
{}from membranes which wind around the circle.  Now, move on the moduli
space of vacua by moving one of the 5-branes around the circle and
bringing it close to the other $N-1$ 5-branes from the other side.  The
strings which were nearly tensionless at the beginning of the process now
have tension of order $L_5$.  However, new strings become nearly
tensionless now.  These originate from membranes which were previous
wound around the circle.  Since they are light, they cannot be
ignored.  They must also be included in our theory, if it has the same
IR behavior as the analogue model.

Note, that unlike the fixed point with (2,0) supersymmetry which is
scale invariant, this theory has a scale, $L_5$.  It is the tension of
strings which are not light near the singularity.  Below this scale
we have the (2,0) field theoretic degrees of freedom and one
scalar is compact.  Above it new degrees of freedom (those which make
the strings) become important.

The effective IR description that we have obtained is similar to that of
\dvv\ and we can identify the fluxes in the same manner and show that
their energies matches the 16 M-theoretic states that form the spinor of
the $SO(5,5,\IZ)$ duality group.  From the M-theory point of view,
the 16 dimensional spinor representation contains 5 momentum modes in
the compact directions, 10 membranes wrapped around 2-cycles of $T^5$
and the completely wrapped transverse 5-brane.
Fluxes which are independent of the compact scalar, $\phi$, exist
already when compactifying on a 4-torus and were discussed in the
previous section.  The rest of the states utilize the compact scalar. As
in \dvv\ the corresponding fluxes are $\partial_0 \int_{\tilde T^5}
\phi$ and $\oint d\phi$ over the five 1-cycles of the
torus.  The $\phi$ dependent terms of the Hamiltonian are
\eqn\hamil{
H = \frac{R}{2\tilde V N} \int d^5 \sigma {\Pi_\phi}^2 +
\frac{1}{2R\tilde V N}
\int d^5 \sigma \partial_i \phi \partial^i \phi ,}
where $\tilde V$ is the volume of $\tilde T^5$ and
$\Pi_\phi$ is the ``overall U(1)'' momentum conjugate to $\phi$.  Its 
zero mode is quantized,  $\Pi_5^0= \frac{2\pi}{L_5}$. The quantum of
energy is then: 
\eqn\energy{E=\frac{R}{2N} (\Pi_\phi^0)^2 = \frac{(2\pi)^2 R}{2N {L_5}^2}.}
The corresponding mass in spacetime is $M=\frac{2\pi}{L_5}$.
The other fluxes, associated with $\oint d\phi$ lead
to new conserved charges. Denoting the base coordinates by $\sigma_i$,
charged states are of the form: 
\eqn\stateso{ \phi (\sigma_i)= \frac{n_i L_5}{\tilde L_i} \sigma_i .}
The energy of these 5 ``winding modes'' can be computed from
\hamil. This leads to the masses
\eqn\lightconem{M_i=\frac{L_5}{\tilde L_i R}= \frac{L_5 L_i}{(2\pi)^2
{l_p}^3},\ \ i=1,\ldots, 4}
and 
\eqn\winds{
M=\frac{L_5}{\tilde L_5 R}= \frac{L_1 L_2 L_3 L_4 L_5}{ (2\pi)^5 {l_p}^6}.}
The first four states can be identified with membranes wrapped around 2
dimensions of the spacetime $T^5$ and the fifth corresponds to the
completely wrapped transverse 5-brane.

Since, as we discussed above, our theory has finite tension strings,
it is tempting to interpret the U-duality group $SO(5,5,\IZ)$ of the
theory as the T-duality of these strings.  This interpretation can be
carried further. M-theory on $T^5$ has one dimensionful parameter,
$l_p$, and 25 dimensionless parameters.  15 of them originate from the
metric, and 10 {}from the $C$ field. A string theory on a 5-torus has
the fundamental scale, the string scale (its $\alpha'$), 15
dimensionless metric parameters and 10 dimensionless parameters in the
$B$ field that couples to that string. The $SO(5,5,\IZ)$ T-duality
group acts on these parameters. Perhaps the full theory which
describes M-theory on $T^5$ is the large $N$ limit of some
non-critical string theory.

\bigskip\bigskip
\centerline{\bf Acknowledgments}\nobreak

We would like to thank O. Aharony, T. Banks, D. Berenstein, R. Corrado,
J. Distler, M. Douglas, R. Entin, W. Fischler, S. Shenker and
L. Susskind for useful discussions. This work was supported in part by
DOE grant \#DE-FG02-96ER40559, by the Robert A. Welch Foundation and by
NSF Grant PHY 9511632.

\listrefs
\end